\documentstyle[12pt,aasms4]{article}

\def\sun{_\odot}

\begin{document}

\title{The Molecule-Rich Tail\\of the Peculiar
Galaxy NGC 2782 (Arp 215)}
\author{Beverly J. Smith}
\affil{CASA, University of Colorado, Box 389, Boulder CO 80309}
\author{Curtis Struck}
\affil{Department of Physics and Astronomy, Iowa State University,
Ames IA  50012}
\author{Jeffrey D. P. Kenney}
\affil{Yale University Astronomy Department, New Haven CT  06520-8101}
\author{Shardha Jogee}
\affil{Department of Astronomy, Caltech, Pasadena CA 91125}

\begin{abstract}

We present the first detection 
of a large quantity of molecular gas 
in the extended tail of an
interacting galaxy.
Using the NRAO 12m telescope, we have detected
CO (1 $-$ 0) at five locations in
the eastern tail
of the peculiar starburst galaxy NGC 2782.
The CO velocities and narrow (FWHM $\sim$ 50 km~s$^{-1}$) 
line widths
in these positions agree with
those seen in HI, confirming that the molecular
gas is indeed associated with the tail rather
than the main disk. 
As noted previously, the gas in this tail has
an apparent `counter-rotation' compared 
to gas in the core of the galaxy, probably because the tails
do not lie in the same plane as the disk.
Assuming the standard Galactic conversion
N$_{H_2}$/I$_{CO}$
factor, these observations 
indicate a total molecular gas mass of 6 $\times$ 10$^8$ M$\sun$
in
this tail.  
This 
may be an underestimate of
the total H$_2$ mass if the gas is metal-poor.
This molecular gas mass, and the implied H$_2$/HI mass ratio of
0.6, are higher than that
found in many dwarf irregular
galaxies.
Comparison with an available H$\alpha$ map of this galaxy, however,
shows that the rate of star formation in this feature is extremely low
relative to the available molecular gas, 
compared to L$_{H\alpha}$/M$_{H_2}$ values
for both spiral and irregular galaxies.  Thus the timescale
for depletion of the gas in this feature is very long.

\end{abstract}

\keywords{Galaxies: Individual (NGC 2782) $-$ Galaxies: ISM 
$-$ Galaxies: Interactions}

\section{Introduction}

Extragalactic tails and bridges
are the best-known signatures
of recent gravitational interactions between galaxies
(e.g., Toomre $\&$ Toomre \markcite{tt72}1972).
They are also often the birthsites of
stars (Schweizer \markcite{s78}1978; Mirabel et al. \markcite{m91}1991, 
\markcite{m92}1992).
Understanding when and how star formation is initiated
in unusual environments such as tails and bridges
can provide important clues to the processes
governing star formation in general, while studying
the properties of gas in tails and bridges
may provide information about gas behavior during
galaxy interactions and collisions.

One poorly-understood factor in star formation
initiation in tails and bridges
is the 
amount and distribution of 
molecular gas.  
Because molecular gas is the material out of which stars
form, it is important to measure its distribution and
mass in these structures, 
to test theories of how star formation is triggered in tails and bridges,
and to understand gas phase changes during galaxy collisions
and tail/bridge formation.
In an earlier NRAO 12m telescope survey of tidal tails in six
interacting/merging systems, we
searched for CO emission with no success (Smith $\&$ Higdon 
\markcite{sh94}1994).
CO was also not found in the tail of the Leo Triplet galaxy
NGC 3628 (Young 
et al. \markcite{y83}1983), the tidal dwarf in the Arp 105 
system (Duc $\&$ Mirabel \markcite{dm94}1994;
Smith et al. \markcite{smith98}1998), or the HI-rich bridge
of NGC 7714/5 (Struck et al. \markcite{str98}1998).
The only locations where CO has been found outside of  
the main disk of a galaxy is 
a small concentration 
($\sim$10$^6$ M$\sun$)
of molecular
gas 
near
an extended arm in the M81 system
(Brouillet et al. \markcite{b92}1992) and a larger mass (10$^9$ M$\sun$)
near the peculiar Virgo Cluster
galaxy NGC 4438 (Combes et al. \markcite{c88}1988).  In the latter case,
this gas is believed to have been removed from the disk
by ram pressure stripping during a high velocity collision with 
its apparent companion, the S0 galaxy NGC 4435 (Kenney et al. 
\markcite{k95}1995).
In this paper, we present the first detection
of a large quantity of molecular gas in an extragalactic tail, the eastern
tail of the peculiar starburst galaxy NGC 2782.

\section{NGC 2782}

The peculiar galaxy NGC 2782 (Figure 1) is an isolated galaxy
with two prominent tails (Smith \markcite{s91}1991, 
\markcite{s94}1994; Sandage $\&$ Bedke \markcite{sb94}1994).
The longer western tail is rich in HI but faint in the optical,
and the atomic gas extends well beyond the observed stars
(Smith \markcite{s91}1991; Jogee et al. \markcite{j98a}1998).
The eastern tail has a gas-deficient optical knot at the tip
(Smith \markcite{s94}1994).
The HI in this tail is concentrated at the base of the stellar
tail (Smith \markcite{s94}1994).
HII regions are visible 
in this location in the Arp Atlas \markcite{a66}(1966) photograph
as well as
the H$\alpha$ map of Evans et al. \markcite{e96}(1996), and
have been confirmed spectroscopically by
Yoshida, Taniguchi, $\&$ Mirayama \markcite{ytm94}(1994).

Unlike most known double-tailed merger remnants (e.g., NGC 7252; Schweizer
\markcite{s82}1982), the
main body of NGC 2782 has a exponential light distribution 
(Smith \markcite{s94}1994; Jogee et al. \markcite{j99}1999),
indicating that its
disk survived the encounter that created the tails.
The HI velocities of the NGC 2782 tails (Smith \markcite{s94}1994)
are opposite those expected from the H$\alpha$
(Boer et al. \markcite{b92}1992),
HI (Smith 
\markcite{s94}1994), and CO (Jogee et al. \markcite{j98b}1999)
velocity fields of
the galaxy core, indicating that the tails are probably in a different
plane than the inner disk.
The center of NGC 2782 contains a well-known nuclear starburst
(Sakka et al. \markcite{s73}1973; Balzano \markcite{b83}1983; 
Kinney et al. \markcite{k84}1984),
with an energetic outflow (Boer et al. \markcite{b92}1992; Jogee et al.
\markcite{j88a}1998; Yoshida et al. \markcite{y98}1998). 
The disk of NGC 2782 appears
somewhat disturbed, with three prominent optical `ripples'
(Arp \markcite{a66}1966; Smith \markcite{s94}1994), one of which contains bright
H~II regions
(Hodge $\&$ Kennicutt \markcite{hk83}1983; Smith \markcite{s94}1994; 
Evans et al. \markcite{e96}1996; Jogee et al. \markcite{j98a}1998).

The lack of an obvious companion galaxy to NGC 2782 (Smith \markcite{s91}1991),
as well the presence of two oppositely directed tidal tails, 
suggests that it may be the remnant of a merger.
The survival of the NGC 2782 disk, however, indicates that if it
is a merger, the intruder was probably not of equal mass (Smith
\markcite{s94}1994).
It is possible that the optical concentration at the
end of the eastern tail is the remains of a low mass companion,
connected to the main galaxy by a stellar bridge.
The presence of the ripples in the disk as
well as the lack of HI at the tip of the eastern tail are consistent
with the
hypothesis that this companion passed through the disk
of the main galaxy (Smith \markcite{s94}1994).
The striking gas/star offset in this tail may be an example
of the differing behaviour of gas and stars during a galaxy
collision: the gas may have been left behind 
as
the companion passed through the larger galaxy.
In the longer western tail, in contrast to the eastern tail,
the HI extends beyond the optical tail (Figure 1).
In the above collision scenario, the longer western tail is
material
torn from the main galaxy's outer disk, which may have initially
been more extended in gas than in stars.

In our previous CO survey of tidal features, we searched
for CO in the longer western tail of NGC 2782 with no success
(Smith $\&$ Higdon \markcite{sh94}1994).   In this paper, we present
new CO observations of 
the shorter eastern tail that
reveal a large quantity of molecular gas in this feature.
As noted above, this feature may be either a tail
or a bridge plus companion, depending on how it formed.
For convenience throughout this paper, we will simply refer
to it as a tail. However, we note that it has
some morphological differences from `classical' tidal
tails (e.g., the Antennae), in particular, the concentration
of gas at the base of the stellar feature is unusual.
Throughout this paper, we assume a distance of 34 Mpc
(H$_o$ = 75 km s$^{-1}$ Mpc$^{-1}$) to NGC 2782.

\section{Single Dish CO (1 $-$ 0) Observations}

NGC 2782 was observed in the 
CO (1 $-$ 0) line during
1996 December, 
1997 April, May, and October,
and 1998 October
using the 3mm SIS
receiver on the NRAO 12m telescope.
Two 256$\times$2 MHz filterbanks, one for each
polarization, were used for these observations,
providing a total bandpass of 1300 km s$^{-1}$
centered at 2555 km s$^{-1}$ with a
spectral resolution of 5.2 km s$^{-1}$.
A nutating subreflector with a beam throw of 3$'$
was used, and each scan was 6 minutes
long. The beamsize FWHM is 55$''$ at 115 GHz.
The pointing was checked periodically with
bright continuum sources and was consistent
to 10$''$.  The system temperatures ranged
from 300 to 400 K.  
Calibration was accomplished using an ambient chopper wheel.

We observed 17 positions in the NGC 2782 system.  Fifteen of these
were 
arranged in 
a 5 $\times$ 3 grid at 25$''$ spacing.  These include the center and
8 surrounding positions, as well as six positions
in the eastern tidal tail.
In addition, we re-observed the position in the western
tidal tail previously observed by Smith $\&$ Higdon \markcite{sh94}(1994)
and observed another position at the tip of the western tail.
In Figure 1,
these positions are marked on the HI and optical
maps from Smith \markcite{s94}(1994) and Jogee et al. 
\markcite{j98a}(1998).

We have detected CO emission at 14 of the 17 observed positions in NGC 2782:
the center, the surrounding positions, 
and five
positions in the eastern tail.
The sixth position in the eastern
tail was not detected.
The position in the western tail, previously observed but
undetected by Smith $\&$ Higdon \markcite{sh94}(1994), remains undetected.
The position at the tip of the western tail was also undetected.
The final summed scans 
are shown in Figures 2 $-$ 4. Integrated fluxes,
rms noise levels, peak velocities, and line widths
are provided in Table 1.
For the first position in the western tail, the new data
have been combined with the older data.
Note that the noise levels for the positions in 
the tails are considerably lower than for the other positions.

\section{Molecular Gas in the Eastern Tail of NGC 2782}  

The most striking result of our 12m observations is
the detection of CO out in the eastern HI structure.
The central velocities of the CO lines in that tail and
the narrow CO line widths 
are consistent with those in HI (Smith \markcite{s91}1991, \markcite{s94}1994),
showing that this molecular gas is associated with
the tail rather than the main disk.
This is illustrated in Figure 5, where velocity is plotted against right 
ascension for both the 21 cm HI data from Smith (1994)
and the new CO data.
The CO and HI in the eastern tail are at a velocity of $\sim$2620 km s$^{-1}$,
redshifted relative
to the systemic velocity of 2555 km s$^{-1}$.  The HI in the western tail is blueshifted.
Both the CO and HI gas in the disk, however, are blueshifted to the east of the nucleus
and redshifted to the west of the nucleus.
The molecular gas in the eastern tail, like the HI,
shows a reversal in velocity, an apparent `counter-rotation',
with respect to the gas in the inner disk.
As noted previously (Smith 1994), the tails
are probably not in the same plane as the disk so this may
not represent a true counter-rotation in the same plane.

Converting the CO fluxes for this tail into molecular gas masses is
very uncertain, because of
possible metallicity and CO self-shielding effects.
In tidal features, where the column densities and metallicities
tend to be low, the 
Galactic 
I$_{CO}$/N$_{H_2}$ 
ratio may underestimate the amount of
molecular gas (Smith $\&$ Higdon \markcite{sh94}1994) (see Section 7).
The gas in the eastern tail of NGC 2782
may be metal-poor
(Yoshida et al. \markcite{y94}1994),
although this has not yet been
quantified. 
Using the Galactic 
N$_{H_2}$/I$_{CO}$ 
ratio 
for this tail may therefore underestimate the total amount of
molecular gas present in it. 
For convenience in comparing with other galaxies and other tails/bridges,
we
will use this conversion factor
(2.8 $\times$ 10$^{20}$ cm$^{-2}$/(K km s$^{-1}$); Bloemen et al. 
\markcite{b86}1986),
with the understanding
that the H$_2$ mass it provides may be a lower limit to the 
true molecular gas mass in this feature.
Possible variations to this conversion factor
are discussed in detail in Section 7.

Assuming the emission fills the beam (coupling efficiency $\eta$$_c$ = 0.82),
the Galactic conversion factor gives an average H$_2$ column density
for the five observed locations
in the eastern tail of
2 $\times$ 10$^{20}$ cm$^{-2}$.
The CO flux does not vary wildly from position to position
in this tail, showing that molecular clouds are distributed throughout
the feature, not concentrated in a single location.
Integrating over all five positions in the eastern tail,
the Galactic 
N$_{H_2}$/I$_{CO}$ value gives
a total molecular gas mass for this tail 
of 6 $\times$ 10$^8$ M$_{\sun}$.
This is the first detection of such a large quantity of molecular
gas in a tail or bridge.
This mass is two
orders of magnitude higher than that in the possible M81 cloud
(Brouillet et al. \markcite{b92}1992), and is similar to or greater than 
that found in irregular galaxies using the same conversion
factor (Combes \markcite{c85}1985; Tacconi
$\&$ Young \markcite{ty87}1987).

The molecular to atomic gas mass ratio
for this tail is thus
0.6.  This is higher than 
the M$_{H_2}$/M$_{HI}$ ratio
derived for most dwarf irregular galaxies with the same
conversion factor (Combes \markcite{c85}1985; Tacconi
$\&$ Young \markcite{ty87}1987; Israel, Tacconi, $\&$ Baas 
\markcite{itb95}1995).
It is also higher than that found
for Scd and Sm
galaxies,
but lower than the global values for earlier 
high mass spiral galaxies (Young
$\&$ Knezek \markcite{yk89}1989).  This ratio is consistent with
the value found in the outer regions of
the Milky Way and other spiral galaxies, at galactic radii of 5 $-$ 15 kpc (Bloemen
et al. \markcite{b86}1986; 
Tacconi $\&$ Young \markcite{ty86}1986; Kenney, Scoville,
$\&$ Wilson \markcite{ksw91}1991).  

\section{Star Formation in the Eastern Tail of NGC 2782}

To investigate the processes that trigger star formation
in tails and bridges, 
it is important
to quantify the star formation rates, efficiencies, and morphologies
in these structures.
In Figure 6, we compare the HI structure of the eastern tail 
with the H$\alpha$ map from Jogee et al. \markcite{j98a}(1998).
This map shows at least nine H~II regions in this tail
(Table 2).
Four of these were previously tabulated by Evans et al. 
\markcite{e96}(1996).
Star formation is well-distributed throughout
this feature, not concentrated
in a single location.
Calibrating the H$\alpha$ image using the total H$\alpha$ flux for NGC 2782
from Smith \markcite{s94}(1994) gives a total H$\alpha$ luminosity for
this tail of 4.0 $\pm$ 1.7 $\times$ 10$^{39}$ L$\sun$.
This 
falls within the range
spanned by irregular galaxies
(Hunter $\&$ Gallagher \markcite{hg85}1985;
Hunter, Hawley, $\&$ Gallagher \markcite{hhg93}1993), and is 
close to the observed L$_{H\alpha}$
for well-known irregular galaxy NGC 6822 (Hunter et al. 
\markcite{hhg93}1993).
Assuming an extended Miller-Scalo initial mass function (Kennicutt 
\markcite{k83}1983) and no extinction correction, 
the total star formation rate for the eastern NGC 2782 tail is 
therefore between 0.01 and 0.05 M$\sun$ year$^{-1}$.
The H$\alpha$ luminosities for the observed H~II regions in this
tail range from 5 $\times$ 10$^{37}$ erg s$^{-1}$ to 
3 $\times$ 10$^{38}$ erg s$^{-1}$ (Table 2),
with the more luminous regions being in the south.
These luminosities are similar to those of the
brightest H~II regions in NGC 6822 (Hodge, Lee, $\&$ Kennicutt
\markcite{hlk89}1989), thus these H~II regions are not extremely luminous, being more than
an order of magnitude fainter than the 30 Doradus H~II region in the Large
Magellanic Cloud (Faulkner \markcite{f67}1967; Kennicutt \markcite{kennicutt84}1984).

The ratio of the star formation rate to the available molecular
gas for this feature,
L$_{H\alpha}$/M$_{H_2}$, is
0.002 L$\sun$/M$\sun$. This is very low, 
compared to 
global values for high mass galaxies 
(0.001 $-$ 1 L$\sun$/M$\sun$, with the majority between 0.0025 and 0.1 L$\sun$/M$\sun$;
Young et al. \markcite{y96}1996).
This implies that the timescale
for depletion of the available gas 
by star formation is very long, about 20 billion years.
A greater-than-Galactic N$_{H_2}$/I$_{CO}$ ratio in the NGC 2782 
tail would make
these differences even more extreme.
The H$\alpha$/CO ratio for this tail is also low compared to 
irregular galaxies.   For 
eight irregular galaxies with CO and H$\alpha$ measurements
available (from Tacconi $\&$ Young \markcite{ty83}1983; 
Hunter $\&$ Gallagher \markcite{hg96}1996; Hunter et al. 
\markcite{hhg93}1993; 
Young et al. \markcite{y96}1996; Madden et al. 
\markcite{m97}1997; Israel \markcite{i97}1997), the L$_{H\alpha}$/M$_{H_2}$ ratios
(using the Galactic conversion factor for comparison purposes) range from 
$\ge$0.01 to 1.9, much higher than our value for the NGC 2782 eastern tail.
Therefore
this tail has a very low star formation rate relative to its CO flux,
compared to global values for galaxies in general.

\section{Comparison With Other Tail/Bridge Features}

In Table 3, we compare the HI and implied H$_2$
column densities 
of the eastern NGC 2782 tail with five
other extended features: the star-forming tail in the Antennae galaxy (NGC 4038/9),
the NGC 7714/5 bridge, the molecular gas concentration
outside of the main disk of NGC 4438, the Magellanic Irregular
in the Arp 105 system, and the northern tail of NGC 4676 (the `Mice').
We also include derived M$_{H_2}$/M$_{HI}$ values in this table.
As before, we are assuming the Galactic 
N$_{H_2}$/I$_{CO}$
conversion factor 
for convenience; in Section 7, we discuss possible variations in this
factor.
For the star-forming Antennae tail and the Arp 105 irregular,
the 
M$_{H_2}$/M$_{HI}$ 
upper limits in the CO beam
are $\le$0.2, much less than the detected level in
the eastern NGC 2782 tail.  
In the bridge of the interacting pair NGC 7714/5,
the CO/HI upper limit is even lower,
implying
M$_{H_2}$/M$_{HI}$ $\le$ 0.06.
On the other hand, in the NGC 4438 source, the CO mass is large
compared to the HI mass 
(M$_{H_2}$/M$_{HI}$ $\sim$ 5).  
Thus there is a wide range in the CO/HI
ratios in these features.
For NGC 4676 and the other tails measured but not detected in CO
(Young et al. \markcite{y83}1983; Smith $\&$ Higdon 
\markcite{sh94}1994; Duc $\&$ Mirabel \markcite{dm94}1994;
Duc et al. \markcite{d97}1997), including our new measurements of the longer western tail 
of NGC 2782, 
the HI column densities are less than that in the eastern NGC 2782 tail,
and the derived
M$_{H_2}$/M$_{HI}$ upper limits in the CO beam are similar to or
higher than
the ratio for the NGC 2782 eastern tail,
so we are not able to make any 
strong comparisons.

In Table 4, we compare 
the star forming properties of 
the eastern NGC 2782 tail with those of 
the other objects in Table 3.
For the NGC 4438 feature, much of the ionized gas 
may have been ionized by shocks rather than young stars (Kenney
et al. \markcite{k95}1995). Therefore, in Table 4 we list
the observed 
H$\alpha$ luminosity as an upper limit
to the H$\alpha$ luminosity from young stars.
Arp 105 is not included in Table 4,
although star formation is
on-going in the Arp 105 structure
(Duc $\&$ Mirabel \markcite{dm94}1994),
because no global H$\alpha$ flux has been published for this
feature to date.

The H$\alpha$ luminosities are similar for 
the NGC 2782 tail, the NGC 7714/5 bridge, and the Antennae
tail and a few times larger for the NGC 4676 tail.  
For NGC 4438, the upper limit to L$_{H\alpha}$ is similar to the
measured values for the other systems.
The CO luminosity for the eastern NGC 2782 tail 
and the NGC 4438 clump
are much higher than in the other systems, and therefore the implied 
L$_{H\alpha}$/M$_{H_2}$ 
ratios are much lower, if the 
N$_{H_2}$/I$_{CO}$
ratios are similar.  
The L$_{H\alpha}$/M$_{H_2}$ ratio of NGC 2782, 
and the upper limit for NGC 4438, 
are more than 7 times lower than the lower limit for the Antennae dwarf
and 3 $-$ 4 times lower than the lower limits for 
the NGC 7714/5 and NGC 4676 features.
Thus either the rate of star formation per molecular gas
mass differs from system to system, being lowest in NGC 2782 and NGC 4438,
or the 
N$_{H_2}$/I$_{CO}$ ratios are lower in NGC 2782 and NGC 4438 than in the other
objects.   These possibilities are discussed in Sections 7 $-$ 9.

We also note that the spatial distributions 
of the H~II regions
vary from feature to feature for the objects in Table 4, so
the average ambient ultraviolet flux differs from
object to object.
The nine H~II regions in the NGC 2782 tail are spread out
over
a total area of 60 kpc$^2$, while in the Antennae, the three
H~II regions found by 
Mirabel et al. \markcite{m92}(1992) are located within
a 10 kpc$^2$ region.
In the NGC 7714/5 bridge, 
the area subtended
by the star forming regions is $\sim$36 kpc$^2$, 
while in the NGC 4676 tail, the H~II regions in the 55$''$ (23 kpc) CO beam are aligned along
a narrow ridge $\sim$2 kpc or less in width (\markcite{h95}Hibbard 1995; \markcite{hg96}Hibbard
$\&$ van Gorkom 1996).
Thus the NGC 2782 H~II regions are more spread out than the HII regions 
in these other features, 
therefore the ambient UV field is weaker.

The 
H$\alpha$ luminosity functions
also
appear to differ from feature to feature.
Most of the observed
H$\alpha$ in the Antennae dwarf is arising from a single 
luminous H~II region of 1.4 $\times$ 10$^{39}$ erg s$^{-1}$;
the NGC 4676 tail contains several knots with similar luminosities
(\markcite{h95}Hibbard 1995).
These H~II regions are 
more luminous than any individual H~II region in the NGC 2782 feature (Table 2).
In NGC 7714/5, the three
brightest H~II regions in the bridge
(Gonz\'alez-Delgado et al. \markcite{g95}1995) 
are also more luminous than any in the NGC 2782 tail,
but less luminous than the brightest in the Antennae tail. 

\section{Possible Variations in the N$_{H_2}$/I$_{CO}$ Ratio}

In comparing the NGC 2782 tail to other features, 
we must first address the question of possible system-to-system differences
in the 
N$_{H_2}$/I$_{CO}$ ratio.   The parameters that may
affect this ratio include the metallicity and dust content,
the column and volume density, and the ambient ultraviolet
radiation field.

Low dust extinction,
as well as low C and O abundances,
leads to more CO destruction and therefore
smaller CO cores in low metallicity molecular clouds (Maloney $\&$ Black
\markcite{mb88}1988;
Maloney \markcite{m90}1990; Maloney $\&$ Wolfire \markcite{mw96}1996).
CO interferometric studies of nearby dwarf galaxies support
this scenario;
the virial masses implied by the linewidths are often higher than 
H$_2$ masses derived from CO fluxes using the standard Galactic
conversion ratio
(Dettmar $\&$ Heithausen
\markcite{dh89}1989; 
Rubio et al. \markcite{r91}1991, \markcite{r93}1993a,\markcite{rlb93b}b;
Wilson \markcite{w94}1994, \markcite{w95}1995; 
Arimoto et al. \markcite{a96}1996).
There is some suggestion that the 
N$_{H_2}$/I$_{CO}$ ratio in low metallicity systems
scales with abundance, but with large scatter (Wilson \markcite{w95}1995;
Arimoto et al. \markcite{a96}1996).
One of the reasons for the 
observed
system-to-system CO/HI variations seen in Table 3 may therefore be 
abundance
variations.  At this point, however, not enough information is available
about these features to test this hypothesis.
Less-than-Galactic oxygen abundances of 12 + log[O/H] = 8.4 and 8.6 have been
derived for
the Antennae and Arp 105 features, respectively
(Mirabel et al. \markcite{m92}1992; Duc $\&$ Mirabel \markcite{dm94}1994).
These are similar to the metallicity of the Large Magellanic Cloud (Dufour
\markcite{d84}1994;
Russell $\&$ Dopita
\markcite{rd90}1990), and lower than the average
value for the Milky Way (12 + log[O/H] = 9.0; Shaver
et al. \markcite{s83}1983).
For the Large Magellanic Cloud, an enhanced 
N$_{H_2}$/I$_{CO}$ has been inferred
(Cohen \markcite{c88}1988; Israel $\&$ de Graauw \markcite{id91}1991; Mochizuki
et al. \markcite{m94}1994; Poglitsch et al. \markcite{p95}1995).

For the other four objects in Table 3, abundance
analyses have not yet been undertaken. 
For the eastern tail of NGC 2782, the H~II region studied
by Yoshida et al. \markcite{y94}(1994) shows an enhanced [O~III] $\lambda$5007/H$\beta$
ratio, hinting at a less-than-solar metallicity, however, this
has not yet been quantified.
The nucleus of NGC 7714 has been shown to be metal-poor 
(French \markcite{f80}1980; Garc\'ia-Vargas et al. \markcite{g97}1997), but 
at present no abundance study has been done for
the gas in the NGC 7714/5 bridge, the NGC 4438 clump, or the NGC 4676 tail.
Thus it is not yet possible to determine how metallicity
may be affecting the 
N$_{H_2}$/I$_{CO}$
fluxes in these features.

CO and H$_2$ self-shielding variations may also produce the observed CO/HI
differences in these features.
CO becomes self-shielding at higher column densities
than H$_2$,
leading to higher
N$_{H_2}$/I$_{CO}$
ratios at column densities N$_H$ $\le$ 10$^{21}$ cm$^{-2}$
(van Dishoeck $\&$ Black \markcite{vb88}1988;
Lada et al. \markcite{l88}1988; Blitz,  Bazell, and D\'esert 
\markcite{bbd90}1990).
At column densities N$_H$ $\sim$ 5 $\times$ 10$^{20}$ cm$^{-2}$ or lower,
H$_2$ self-shielding also becomes an issue.
In the local interstellar medium, 
the proportion of gas in molecular form 
decreases rapidly 
at a threshold level of $\sim$5 $\times$ 10$^{20}$ cm$^{-2}$
(Savage et al. \markcite{s77}1977; Federman et al. \markcite{f79}1979).
This threshold increases with decreasing density and
metallicity (Elmegreen \markcite{e89}1989).

The column densities of the features in Table 3 are 
in the range where the lack of CO and possibly H$_2$ self-shielding may
be important.
However, the CO/HI ratio is not correlated with HI + H$_2$ column
density in this small sample.
For the NGC 4438, NGC 2782, and Antennae structures, there is a 
trend of decreasing CO/HI ratios with decreasing HI + H$_2$ column densities.
In NGC 4438, it appears as if both the CO and H$_2$ thresholds are 
exceeded, and the CO/HI ratio is very high.
In the Antennae galaxy, in contrast, CO (and maybe H$_2$) may not be well-shielded,
and the CO/HI ratio is very low.  NGC 2782 lies between these two
extremes.
Our detection of CO in this tail implies that the CO
threshold is exceeded in at least portions of tail, but maybe
not over the entire feature.
NGC 7714/5 and Arp 105, however, do not fit this trend, while for
NGC 4676, the CO/HI upper limit is too high to be able to make any strong
constraints.
Arp 105 has a similar HI column density to NGC 2782 but less CO.
NGC 7714/5 has 
an HI column density higher than the expected CO threshold, and higher
than that of NGC 2782, and yet
has a very low CO/HI upper limit.  Perhaps this bridge has
significantly lower abundances than the other features,
and so a higher CO self-shielding threshold.  

A related reason for the 
observed CO/HI differences in Table 3
may be variations in the clumpiness 
of the gas within the CO beam.
In NGC 4438, the physical size
of the CO
beam is only 2.2 kpc, 2.7 $-$ 4.5 $\times$ less than in
the other systems, so a higher average H$_2$ and CO column density is 
not surprising.  
In Arp 105 and NGC 4676, the beamsizes are 30 kpc and 23 kpc, respectively, so lower average CO 
column densities are also not unexpected.
In the Antennae and NGC 7714/5 features, however, the beam subtends
6 kpc and 10 kpc, respectively, compared to 9 kpc in NGC 2782, yet the CO is much fainter.
Thus there is no trend of decreasing CO/HI ratio with beamsize.
Within the beam, however, there may be variations in how the gas is clumped;
perhaps in the NGC 2782 feature, the CO self-shielding limit is exceeded over
larger portions of the feature and so beam-dilution is less of a factor.
This issue could be addressed with higher resolution CO observations.

Another important factor which affects the 
N$_{H_2}$/I$_{CO}$
ratio 
is the ambient ultraviolet flux, which may be higher in
the Antennae, NGC 4676,  and NGC 7714/5 features
than in the eastern NGC 2782 tail (see Section 6).
If two systems have similar low metallicities, dust
contents, and gas densities, the
one with the more intense
ambient
ultraviolet field will have more CO destruction and so a higher average
N$_{H_2}$/I$_{CO}$
ratio
(Maloney $\&$ Wolfire \markcite{mw96}1996).
This is consistent with the difference between the CO/HI ratios of the NGC 2782 tail
and the other features.

We conclude that system-to-system variations in the
N$_{H_2}$/I$_{CO}$ ratio likely play an important role in
determining the CO/HI values of these features. However, without detailed
analyses of the metal abundances in these structures we are not
able to quantify these differences.

\section{Tail/Bridge Formation Mechanisms}

Another factor
which may contribute to the 
observed variations in the CO content of tails/bridges
is 
differences in the formation mechanisms of these structures.
Two distinct processes contribute to
the formation of extragalactic tails and bridges:
tidal forces 
(e.g., Toomre $\&$ Toomre \markcite{tt72}1972)
and 
ram pressure stripping
(e.g., Spitzer $\&$ Baade \markcite{sb51}1951; Struck
\markcite{str97}1997).  
The relative importance of these two mechanisms
probably varies from
system to system: in small impact parameter collisions
between gas-rich galaxies, cloud-cloud impacts and other
hydrodynamical effects may be important in forming
tails/bridges, while features pulled out during
distant encounters may be largely tidal.
Dissipation in the gas, in addition to possible
pre-collision differences in the gaseous and stellar distributions,
may cause large offsets between the gas and
the stars in bridges and tails.
Such offsets have been found in a number of
systems (Wevers et al. \markcite{w84}1984; Smith 
\markcite{s94}1994; Hibbard $\&$ van
Gorkom \markcite{hv96}1996; Smith et al. \markcite{s97}1997).
Gas dissipation and shocks can occur both during the initial encounter
and also
during subsequent passages, as tails fall back into
the main galaxies (e.g., Hibbard and Mihos \markcite{hm95}1995).

The six features in Table 3 are quite different
morphologically.  The Antennae and Arp 105 structures are end-of-tail clumps
(van der Hulst \markcite{v79}1979; van der Hulst et al. \markcite{vmb94}1994;
Duc et al. \markcite{d97}1997),
while in NGC 4676 H~II regions are seen along the full extent of
the tail (\markcite{h95}Hibbard 1995; \markcite{hg96}Hibbard $\&$ van Gorkom 1996).
The NGC 2782 gas is concentrated at the base of
a stellar tail, the targeted region in NGC 7714/5 is 
in the middle of a bridge connecting two galaxies, and the molecular
concentration near NGC 4438 lies out of the plane of the galaxy.
Ram pressure stripping may have played a bigger role
in forming the eastern tail of NGC 2782 and the NGC 4438 clump
than the other features.
The NGC 2782 tail has a big gas/star offset, while
in the Antennae and NGC 4676 tails, the stars and gas are coincident
(van der Hulst et al. \markcite{v94}1994;
Hibbard $\&$ van Gorkom \markcite{96}1996).  In the Arp 105 irregular, the
gas distribution is well-aligned with that of the stars,
except for a slight offset
at the southern end (Duc et al. \markcite{d97}1997).
The NGC 7714/5 bridge is actually two parallel bridges,
one made out of stars, the other of gas, indicating
that both tidal and hydrodynamical forces contributed to the
formation of this feature (Smith
et al. \markcite{ssp97}1997).
In NGC 4438, a distorted optical tail/arm lies 2 kpc away from
the CO concentration (Combes et al. \markcite{c88}1988); it is unclear whether
the molecular gas is associated with this stellar 
structure or not (Kenney et al. \markcite{k95}1995).

We have ordered these six features in terms of the importance
of ram pressure stripping versus tidal forces in creating them.
The ranking is:
1) NGC 4438,
2) NGC 2782, 3) NGC 7714/5, 4) Arp 105, the Antennae, and NGC 4676.
The Antennae and NGC 4676 tails are classical tidal tails
(Toomre $\&$ Toomre \markcite{tt72}1972; Barnes \markcite{b88}1988;
Mihos, Bothun, $\&$ Richstone
\markcite{mbr93}1993);
the Arp 105 structure is also probably tidal (Duc et al. \markcite{d97}1997).
In contrast, the NGC 4438 clump is probably largely a product of ram
pressure stripping (Kenney et al. \markcite{k95}1995).  The observed gas/star offsets in
the NGC 2782 tail and NGC 7714/5 bridge suggest that gas
dissipation played an important role in producing these features,
but the existence of stellar counterparts shows that tidal
forces also contributed (Smith 1994; Smith et al. 1997).

Interestingly, this splash/tidal ranking also correlates with
the CO/HI ratio in these systems.  The two structures
with the most pronounced gas/star morphological differences have the largest
CO abundances.  
One possible explanation for this correlation is simply a metallicity effect.
In NGC 2782 and NGC 4438,
where `splash' probably played an important role and the impact
parameters may have been smaller, the stripped gas, presumably
removed from the inner disk, may be more metal-rich than
material pulled from the outer disk in a grazing tidal interaction.
Therefore the 
N$_{H_2}$/I$_{CO}$ ratio may be lower than in the other systems.
This possibility could be tested with detailed abundance studies.

The observed large CO fluxes from the NGC 4438 and NGC 2782 features
might be considered 
somewhat surprising, in light of theoretical models of molecular
dissociation 
during near head-on collisions.
In fast shocks,
H$_2$ and CO are dissociated (e.g., Hollenbach $\&$ McKee 
\markcite{hm89}1989), so
in an extreme `splash', where high velocity cloud-cloud collisions
occur, 
one may expect a large proportion of the 
molecular gas to be dissociated (e.g., Harwit et al. 
\markcite{h87}1987).
Direct evidence for strong shocks is present in the optical
spectrum of the NGC 4438 CO clump (Kenney et al. \markcite{k95}1995).
The existence of CO in the NGC 2782 and NGC 4438 features proves that, however they
formed, the collisions/encounters were not 
drastic enough to dissociate all of the molecular gas, or, if the
gas were indeed dissociated,
sufficient time has passed for the H$_2$ to reform.
The molecule formation timescale is $\sim$ 10$^9$/n years/cm$^{-3}$
(Hollenbach $\&$ McKee \markcite{hm79}1979),
so assuming the age of the NGC 2782 structure
is $\sim$ 2 $\times$ 10$^8$ years (Smith \markcite{s94}1994),
if the average density in this tail is n $\ge$ 10
cm$^{-3}$, as expected for molecular clouds,
then it is quite possible that the molecules in this feature dissociated
during the collision and now have reformed.

\section{Star Formation Initiation}

One of the surprising results of this study is the low
L$_{H\alpha}$/M$_{H_2}$ ratio for this NGC 2782 tail.
Whether or not star formation is triggered in
a bridge or tail may be due in part to how much
ram pressure stripping versus tidal effect occurred during the encounter.
In a near-head-on encounter, one might expect
more shocks and cloud fragmentation than in a more gentle
tidal encounter. 
Theoretical models suggest that star formation may
be inhibited in `splash' features because of gas heating during
the collision (Struck \markcite{str97}1997), while in tidal features
gravitational compression may enhance star formation 
(Wallin \markcite{w90}1990).
Thus in `splash' features, molecular gas may be distributed in small,
relatively diffuse clouds rather than concentrated in giant molecular
clouds with high column densities.  
These theoretical
results suggest a trend in the star formation rate per molecular
gas mass with
increasing tidal contribution, consistent with our
results:
the two most likely
`splash' candidates have the lowest H$\alpha$/CO ratios
in the group.

A second possibility is that the gas surface density in
the NGC 2782 tail may be below a critical
surface density required for gravitational collapse, as has been surmised
for the outer regions of galactic disks (Kennicutt \markcite{k89}1989).
For a differentially rotating thin gas disk, the critical surface density is
$\Sigma$$_{crit}$ = $\kappa$$\sigma$$_v$/3.36G (Toomre 
\markcite{t64}1964), where
$\kappa$ is the epicyclic frequency, $\sigma$$_v$ is the velocity dispersion,
and G is the gravitational constant.
Although a tail or bridge may not be directly
participating in the overall rotation of a galaxy, 
we apply this argument by assuming that these structures
are thin, and replacing
$\kappa$
by 2$\Delta$v/R, twice the velocity gradient
southwest to northeast 
along the long axis of the feature 
(assuming the southeast to northwest shear in the tail is negligible).

In Table 4, we have included HI velocity gradients and dispersions for
the other features in our sample, as well as their
expected critical surface densities.
For NGC 4438, the HI data from Cayette et al. (1990) are too low S/N to estimate
the velocity gradient and dispersion accurately, so no critical
density is derived.
We note that the quoted velocity gradients are the observed gradients 
along the
features, which do not take viewing perspective into account.
Therefore the derived critical densities are quite uncertain,
perhaps by a factor of a few.
Within these uncertainties, the critical densities of all the features
are similar, and are similar to the observed gas column densities.
For the eastern NGC 2782 tail, the observed N$_H$ is indeed lower 
than the predicted critical density,
suggesting that the gas in this feature may be
relatively stable against gravitational collapse.
In fact, the observed N$_H$ for this tail is {\it greater}
than the expected self-shielding threshold for CO and H$_2$,
and yet {\it less} than the expected 
critical density for gravitational collapse.
This is consistent with the observation of abundant CO with
relatively low H$\alpha$ luminosity.
In the NGC 7714/5 bridge, 
the HI column density alone is so high that
it approaches the critical density for gravitational instability.  The huge 
mass of atomic gas in this bridge alone may be enough to 
ignite the formation
of stars and enhance the efficiency of star formation.
For NGC 4676, 
the derived critical density is higher
than the 
observed gas column density and for the Antennae the 
critical and observed densities are similar, yet these
galaxies have high L$_{H\alpha}$/CO ratios.
This comparison suggests that 
in the NGC 4676 tail
and perhaps in the Antennae feature 
the beam
filling factor may be low and/or the
N$_{H_2}$/I$_{CO}$
ratio may be higher 
compared to the
NGC 2782 tail.

\section{CONCLUSIONS}

Using the NRAO 12m telescope, we have found evidence for 6 $\times$ 10$^8$
M$\sun$ of molecular gas in the eastern tail of NGC 2782. 
Compared to both spiral and irregular galaxies,
the molecular gas content in the eastern tail of NGC 2782 
is very high relative to the current rate of star formation,
implying a very long timescale for gas depletion.
Both the molecular gas and H~II regions in this feature are very extended,
spread out over a total area of 60 kpc$^{-2}$.
Comparison with tidal or `splash' regions in other galaxies 
shows a wide range in CO/HI
and H$\alpha$/CO values.

\vskip 0.2in

We thank the telescope operators and the staff of the NRAO 12m telescope
for their help in making these observations.
We are pleased to acknowledge funding for this project from
a NASA grant 
administered by the American Astronomical Society.
This research has made use of the NASA/IPAC Extragalactic
Database (NED) which is operated by the Jet Propulsion Laboratory
under contract with NASA.

\begin{table}
   {\bf Table 1}\\
   CO (1-0) RESULTS FOR NGC 2782\\ [12pt]
   \begin{tabular}{crrrrrrcrcccclclcrcccc} \tableline
       \multicolumn{1}{c}{Name}&
\multicolumn{6}{c}{Position Observed}
&\multicolumn{1}{c}{T$_R$$^*$ (rms)}
&\multicolumn{1}{c}{$I_{CO}$$^a$}
&\multicolumn{1}{c}{V$_{peak}$}&
\multicolumn{1}{c}{$\Delta$V$^b$}\\
       \multicolumn{1}{c}{}&
\multicolumn{3}{c}{R.A. (1950)}&
\multicolumn{3}{c}{Dec. (1950)}&
\multicolumn{1}{c}{(mK)}& 
\multicolumn{1}{c}{(K km s$^{-1}$)}& 
\multicolumn{1}{c}{(km s$^{-1}$)}&
\multicolumn{1}{c}{(km s$^{-1}$)}&\\
\tableline
Center&9&10&53.6&40&19&15.0&5.1&5.98 $\pm$ 0.22&2560&360\\ 
North&9&10&53.6&40&19&40.0&7.3&3.36 $\pm$ 0.33&2550&400\\ 
South&9&10&53.6&40&18&50.0&6.0&4.37 $\pm$ 0.28&2550&620\\
East&9&10&55.8&40&19&15.0&6.6&2.21 $\pm$ 0.31&2540&420\\ 
West&9&10&51.4&40&19&15.0&6.8&3.37 $\pm$ 0.32&2590&420\\ 
Northwest&9&10&51.4&40&19&40.0&5.1&3.00 $\pm$ 0.27&2570&520\\
Northeast&9&10&55.8&40&19&40.0&6.4&1.87 $\pm$ 0.28&2550&360\\
Southeast&9&10&55.8&40&18&50.0&6.7&2.57 $\pm$ 0.31&2480&420\\
Southwest&9&10&51.4&40&18&50.0&5.0&1.60 $\pm$ 0.24&2640&440\\
North-Far-East&9&10&58.0&40&19&40.0&3.1&0.73 $\pm$ 0.09&2600&180\\
Far-East&9&10&58.0&40&19&15.0&2.6&0.31 $\pm$ 0.06&2620&90\\ 
South-Far-East&9&10&58.0&40&18&50.0&2.3&0.44 $\pm$ 0.05&2630&90\\ 
North-Far-Far-East&9&11&0.2&40&19&40.0&2.5&$\le$0.18&&110$^d$\\
Far-Far-East&9&11&0.2&40&19&15.0&3.1&0.39 $\pm$ 0.07&2620&105\\
South-Far-Far-East&9&11&0.2&40&18&50.0&3.5&0.57 $\pm$ 0.08&2620&110\\
Western Tail I&9&10&41.1&40&20&16.0&3.6$^c$&$\le$0.22$^d$&&80$^d$\\
Western Tail II&9&10&36.1&40&22&22.0&3.1&$\le$0.19$^d$&&80$^d$\\
   \end{tabular}

\footnotesize

\tablenotetext{a}{$I_{CO} = \int{T_R^*dv}$.} 
\tablenotetext{b}{Full width zero maximum (FWZM).}
\tablenotetext{c}{Combined with the 
Smith $\&$ Higdon (1994) data.}
\tablenotetext{d}{Using the HI FWZM line width (Smith 1991).}
\end{table}

\begin{table}
   {\bf Table 2}\\
   The H~II Regions in the Eastern NGC 2782 Tail\\ [12pt]
   \begin{tabular}{cccccccccccccccccc} \tableline
H~II&\multicolumn{3}{c}{R.A. (1950)}&\multicolumn{3}{c}{Dec. (1950)}&L$_{H\alpha}$\\
Region&h&m&s&$\circ$&$'$&$''$&(erg s$^{-1})$\\
\tableline
A&9&10&59.8&40&18&40.3&3.3 $\times$ 10$^{38}$\\
B&9&10&58.9&40&18&46.3&3.1 $\times$ 10$^{38}$\\
C&9&10&60.0&40&19&2.7&1.2 $\times$ 10$^{38}$\\
D&9&11&0.7&40&19&10.7&6.3 $\times$ 10$^{37}$\\
E&9&10&58.7&40&19&14.2&2.2 $\times$ 10$^{38}$\\
F&9&11&1.0&40&19&21.0&1.0 $\times$ 10$^{38}$\\
G&9&10&58.3&40&19&21.6&5.7 $\times$ 10$^{37}$\\
H&9&10&59.5&40&19&23.9&9.7 $\times$ 10$^{37}$\\
I&9&10&58.8&40&19&46.1&4.7 $\times$ 10$^{37}$\\
\tableline
   \end{tabular}
\end{table}

\begin{table}
   {\bf Table 3}\\
   Gas Properties of Selected Tail/Bridge Features$^a$\\ [12pt]
   \begin{tabular}{cccccccccccccccccc} \tableline
Feature&N$_{HI}$&N$_{H_2}$$^b$&N$_{H}$$^b$&M$_{H_2}$/M$_{HI}$$^b$&Notes\\
&(cm$^{-2}$)&
(cm$^{-2}$)&
(cm$^{-2}$)&\\
\tableline
NGC 2782 Eastern Tail&6 $\times$ 10$^{20}$&2 $\times$ 10$^{20}$&
10$^{21}$&0.6&$^c$\\ 
Antennae Tail&4 $\times$ 10$^{20}$&$\le$4.2 $\times$ 10$^{19}$&
4.0 $-$ 4.8 $\times$ 10$^{20}$&$\le$0.2&$^d$\\ 
NGC 7714/5 Bridge&1.6 $\times$ 10$^{21}$&$\le$5.0 $\times$ 10$^{19}$&
1.6 $-$ 1.7 $\times$ 10$^{21}$&$\le$0.06&$^e$\\ 
NGC 4438 CO Clump&9.2 $\times$ 10$^{20}$&2.0 $\times$ 10$^{21}$&
4.9 $\times$ 10$^{21}$&5&$^f$\\ 
Arp 105 Irregular&7 $\times$ 10$^{20}$&$\le$8.0 $\times$ 10$^{19}$&
7 $-$ 9 $\times$ 10$^{20}$&$\le$0.2&$^g$\\ 
NGC 4676 Northern Tail&2.1 $\times$ 10$^{20}$&$\le$5.9 $\times$ 10$^{19}$&
2.1 $-$ 3.2 $\times$ 10$^{20}$& $\le$0.6&$^h$\\
   \end{tabular}

\tablenotetext{a}{All 
values are averaged 
over the 55$''$ 12m beam,
except for NGC 4438, where a 23$''$ beam was used.}
\tablenotetext{b}{The H$_2$ column density and mass were derived using
the Galactic N$_{H_2}$/I$_{CO}$ ratio of 2.8 $\times$ 10$^{20}$
cm$^{-2}$/(K km s$^{-1}$) (Bloemen et al. \markcite{b86}1986),
assuming
the source fills the beam with a coupling efficiency of $\eta$$_c$ = 0.82.
Although this conversion factor may not hold in these features
(see text), we use it for comparative purposes here.}
\tablenotetext{c}{N$_{HI}$
for the NGC 2782 tail was obtained from the Smith \markcite{s94}(1994) 
HI map.  }
\tablenotetext{d}{N$_{HI}$ for the
Antennae dwarf is from van der Hulst et al. \markcite{v94}(1994), as given in
Smith $\&$ Higdon \markcite{sh94}(1994).  
The CO flux for the Antennae dwarf 
is from Smith $\&$ Higdon \markcite{sh94}(1994).}
\tablenotetext{e}{The
HI column density
for the NGC 7714/5 bridge is 
from the Smith et al. \markcite{s97}(1997) 
HI data.
The CO flux is from 
Struck et al. \markcite{str98}(1998).}
\tablenotetext{f}{For the NGC 4438 molecular 
concentration, N$_{HI}$ is from Cayatte
et al. (1990) and the CO flux is from Combes et al. \markcite{c88}(1988).}
\tablenotetext{g}{For Arp 105, the HI column
density is from Duc et al. \markcite{d97}(1997) and the CO upper
limit is from 
Smith
et al. \markcite{smith98}(1998), with the HI FWZM line width
of 220 km s$^{-1}$ from Duc et al. \markcite{1997}(1997).}
\tablenotetext{h}{For NGC 4676, the CO flux is from Smith $\&$ Higdon \markcite{sh94}(1994)
and the HI flux is from Hibbard \markcite{h95}(1995), as quoted in Smith $\&$ Higdon \markcite{sh94}(1994).}

\end{table}

\begin{table}
   {\bf Table 4}\\
   Star Forming Properties of Selected Tail/Bridge Features$^a$\\ [12pt]
   \begin{tabular}{cccccccccccccccccc} \tableline
Galaxy&D$^b$&L$_{H\alpha}$&M$_{H_2}$$^c$&L$_{H\alpha}$/M$_{H_2}$&$\Delta$v/R&$\sigma$$_v$&N$_{crit}$&Notes\\
&(Mpc)&(erg s$^{-1}$)&(M$\sun$)&(L$\sun$/M$\sun$)&(km s$^{-1}$&(km s$^{-1}$)&(cm$^{-2}$)\\
&&&&&kpc$^{-1}$)\\
\tableline
NGC 2782&34&4 $\times$ 10$^{39}$&6 $\times$ 10$^8$&0.002&5.1&20&2 $\times$ 10$^{21}$&$^d$\\
Antennae&23&1.7 $\times$ 10$^{39}$&$\le$2.9 $\times$ 10$^7$&$\ge$0.015&2.2&
10&4 $\times$ 10$^{20}$&$^e$\\
NGC 7714/5&37&1.8 $\times$ 10$^{39}$&$\le$9.4 $\times$ 10$^7$&$\ge$0.005&3.5&10&
6 $\times$ 10$^{20}$&$^f$\\
NGC 4438&16&$\le$1.6 $\times$ 10$^{39}$&8.4 $\times$ 10$^8$&
$\le$0.0005&&&&$^g$\\
NGC 4676&88&1.1 $\times$ 10$^{40}$&$\le$6.0 $\times$ 10$^8$&$\ge$0.01&
5.5&10&9 $\times$ 10$^{20}$&$^h$\\
   \end{tabular}

\tablenotetext{a}{For the Antennae, NGC 7714/5, and NGC 4676,
the tabulated values
are averaged over the 55$''$ NRAO 12m beam.
For NGC 2782 and 4438, 
values for the entire tail are used.}
\tablenotetext{b}{Assuming H$_o$ = 75 km s$^{-1}$ Mpc$^{-1}$, except for the Virgo Cluster galaxy
NGC 4438, where 16 Mpc is used (Jacoby
et al. \markcite{j94}1992; Freedman et al. \markcite{f94}1994).}
\tablenotetext{c}{The H$_2$ masses were derived using
the Galactic N$_{H_2}$/I$_{CO}$ ratio 
(Bloemen et al. \markcite{b86}1986; see discussion in text), 
assuming
$\eta$$_c$ = 0.82.}
\tablenotetext{d}{For NGC 2782,
$\Delta$v/R
and 
$\sigma$$_v$ 
were obtained from the Smith \markcite{s94}(1994) 
HI map.  
The velocity
dispersions in this feature range from 15 $-$ 30 km s$^{-1}$ at 
the base of the tail
to 10 $-$ 20 km s$^{-1}$ in the rest of the feature.}
\tablenotetext{e}{For the Antennae,
$\Delta$v/R and 
$\sigma$$_{v}$ are 
from Hibbard (1998, private communication).
L$_{H\alpha}$ is from Mirabel
et al. \markcite{m92}(1992).}
\tablenotetext{f}{For the NGC 7714/5 bridge, 
$\Delta$v/R
and 
$\sigma$$_v$ 
were obtained 
from the Smith et al. \markcite{s97}(1997) HI data.
L$_{H\alpha}$ is from 
Gonz\'alez-Delgado et al. \markcite{g95}(1995).}

\end{table}
\begin{table}
\tablenotetext{g}{For the NGC 4438 CO clump, the quoted L$_{H\alpha}$ was determined from
the Kenney et al. \markcite{k95}(1995) H$\alpha$+[N~II] map,
assuming
L$_{H\alpha}$ = 0.38(L$_{H\alpha+[N~II]})$, as determined
by optical spectroscopy (Kenney et al.
\markcite{k95}1995).
Because this gas
is shock-excited (Kenney et al. \markcite{k95}1995), this luminosity
is listed as an upper limit;
it is an upper limit to
the H$\alpha$ luminosity due to young stars. }
\tablenotetext{h}{For the NGC 4676 tail, L$_{H\alpha}$ within the 55$''$ 12m beam
was determined 
from the Hibbard $\&$ van Gorkom \markcite{hg96}(1996) map 
by Hibbard (1998, private communication), assuming
L$_{H\alpha}$ = 0.7(L$_{H\alpha+[N~II]}$).
M$_{H_2}$ is from Smith $\&$ Higdon \markcite{sh94}(1994).  $\Delta$v/R 
and $\sigma$$_v$ are from Hibbard $\&$ van Gorkom \markcite{hv96}(1996) map
as given by
Hibbard (1998, private communication).}

\end{table}

{\bf Captions}

\figcaption[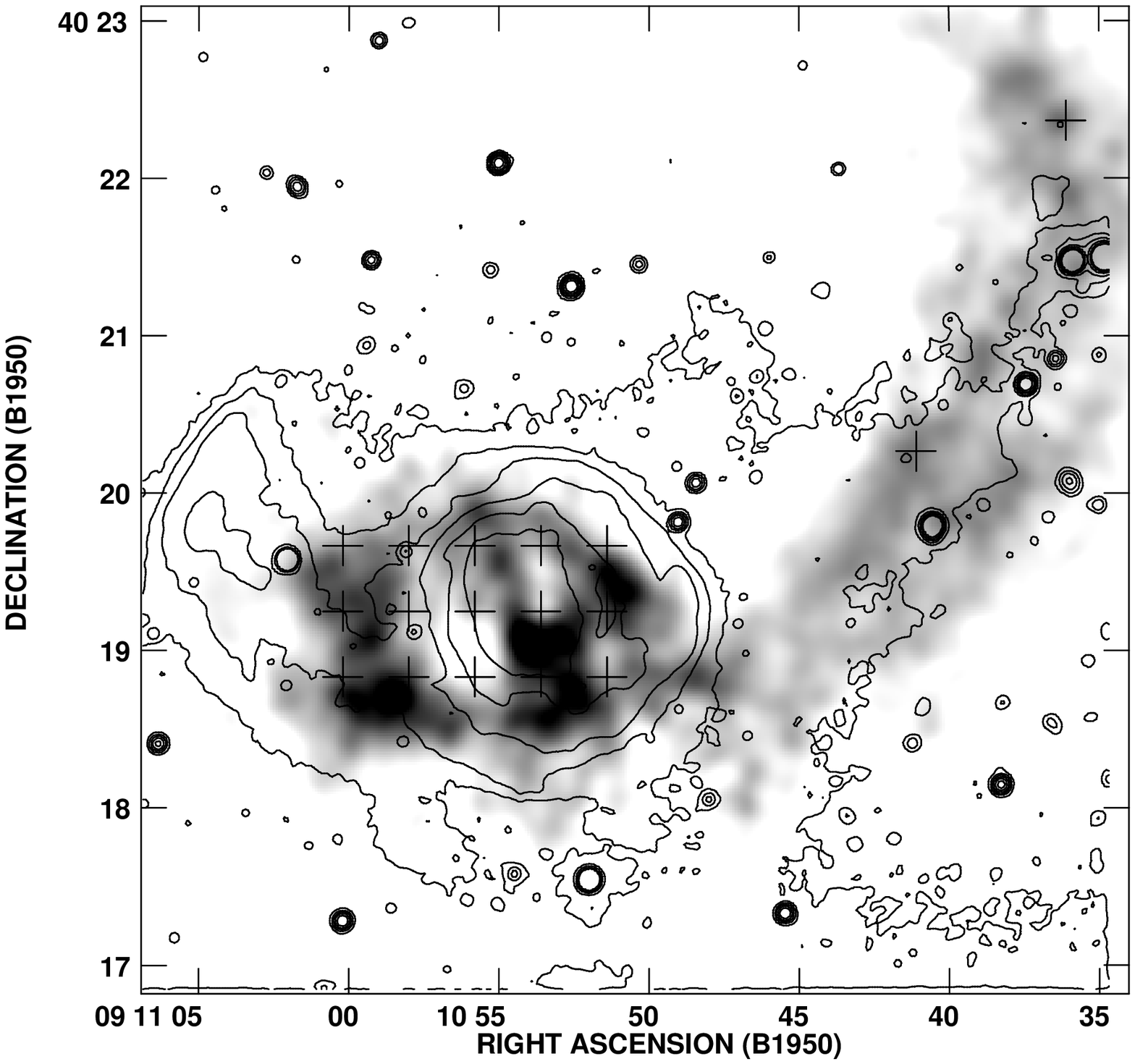]{The 21 cm HI map of NGC 2782 (greyscale; from Smith 1994),
overlaid on the optical B band image (contours; from Jogee et al. 1998).
The optical map has been smoothed to 3$''$ resolution; the HI map,
to 12$''$.
The 17 positions observed in the CO (1 $-$ 0) line
with the NRAO 12m telescope are marked as crosses.}

\figcaption[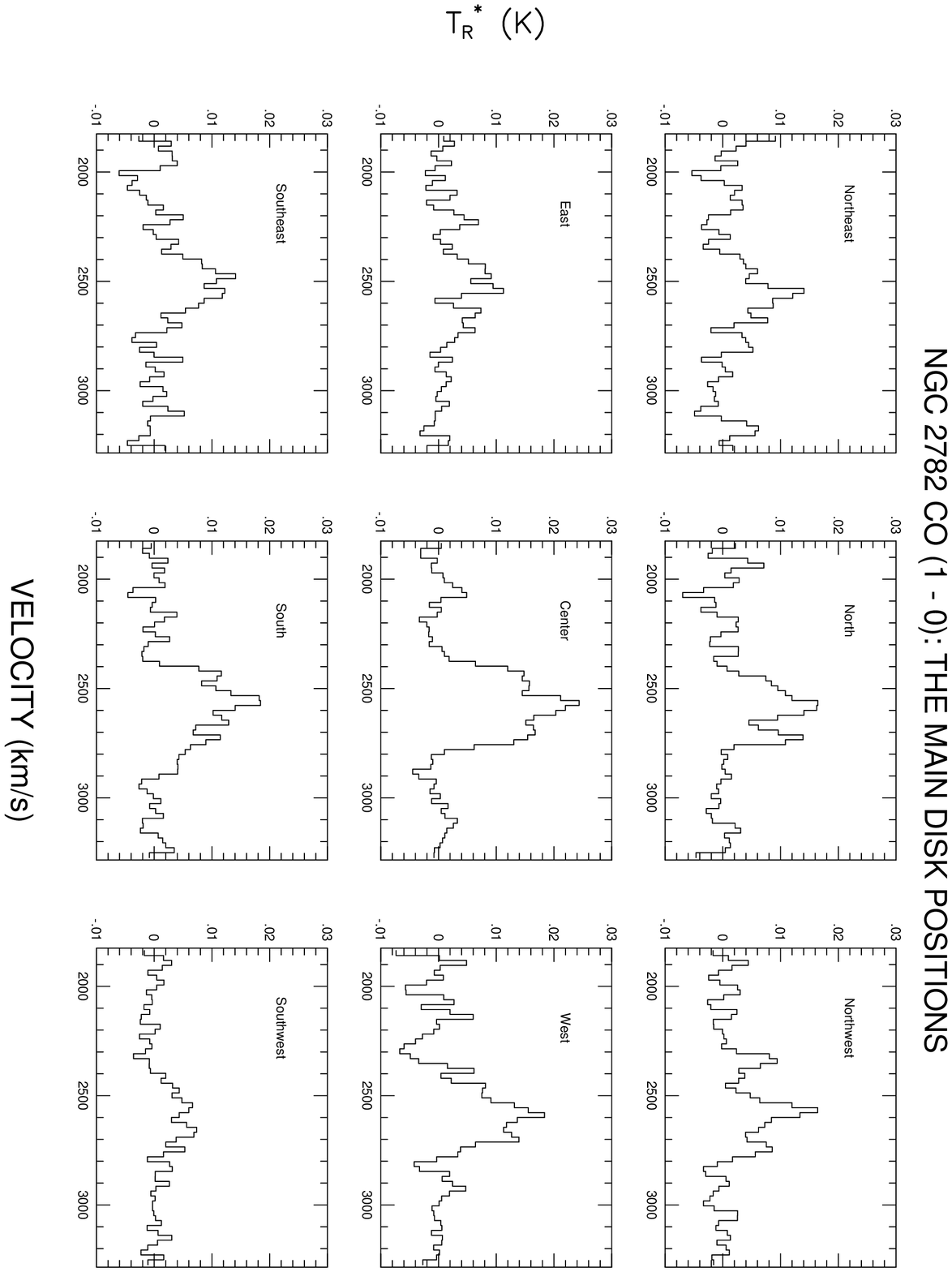]{The CO spectra for the nine positions in the main
disk of NGC 2782: the center and the surrounding eight positions.
For displaying purposes, these spectra have been smoothed by a 36 km
s$^{-1}$ boxcar and then resampled at 21 km s$^{-1}$ spacing.}

\figcaption[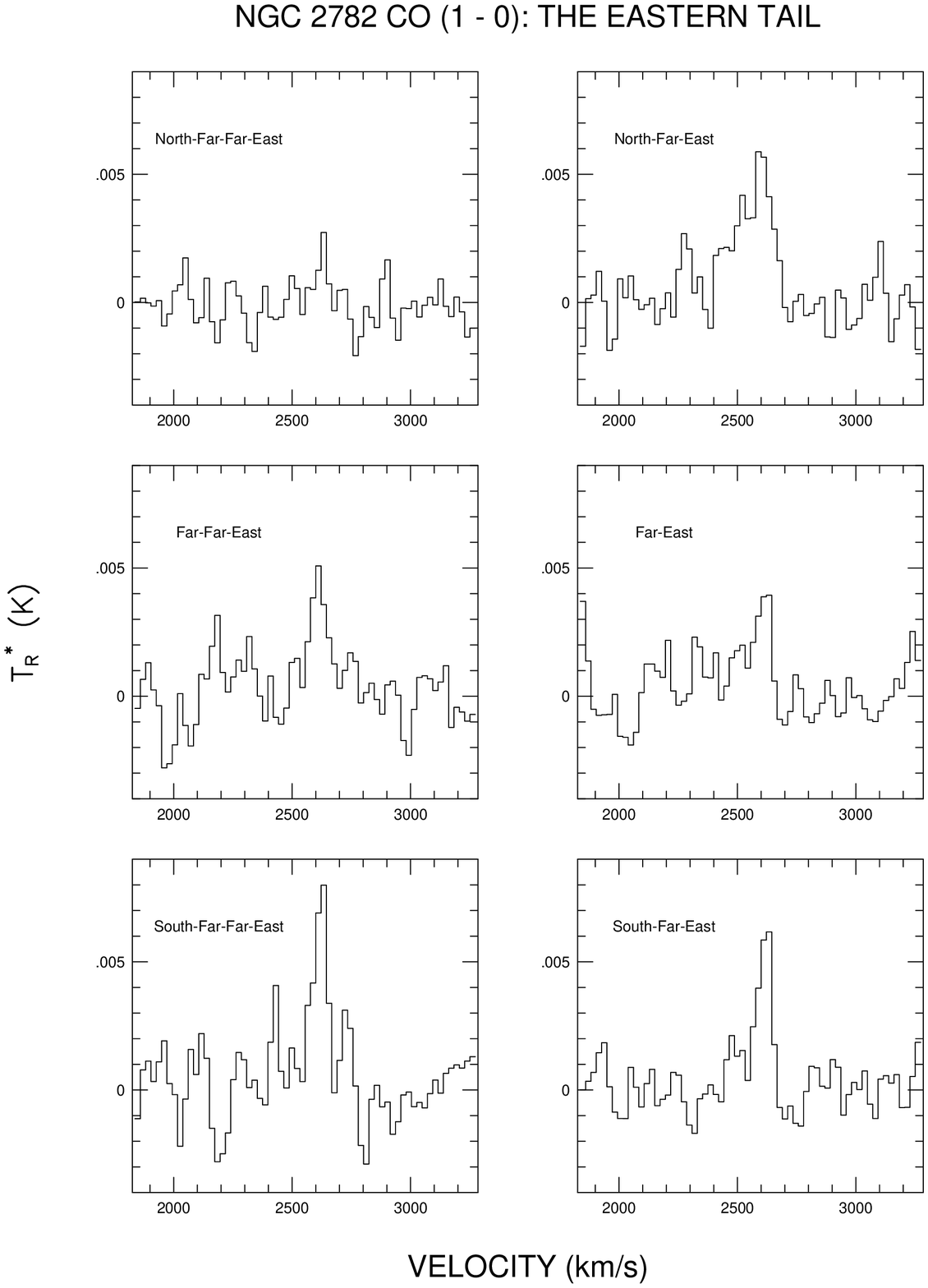]{The CO spectra for the six positions in the eastern
tail of NGC 2782.
These data have been smoothed and resampled as in Figure 2.}

\figcaption[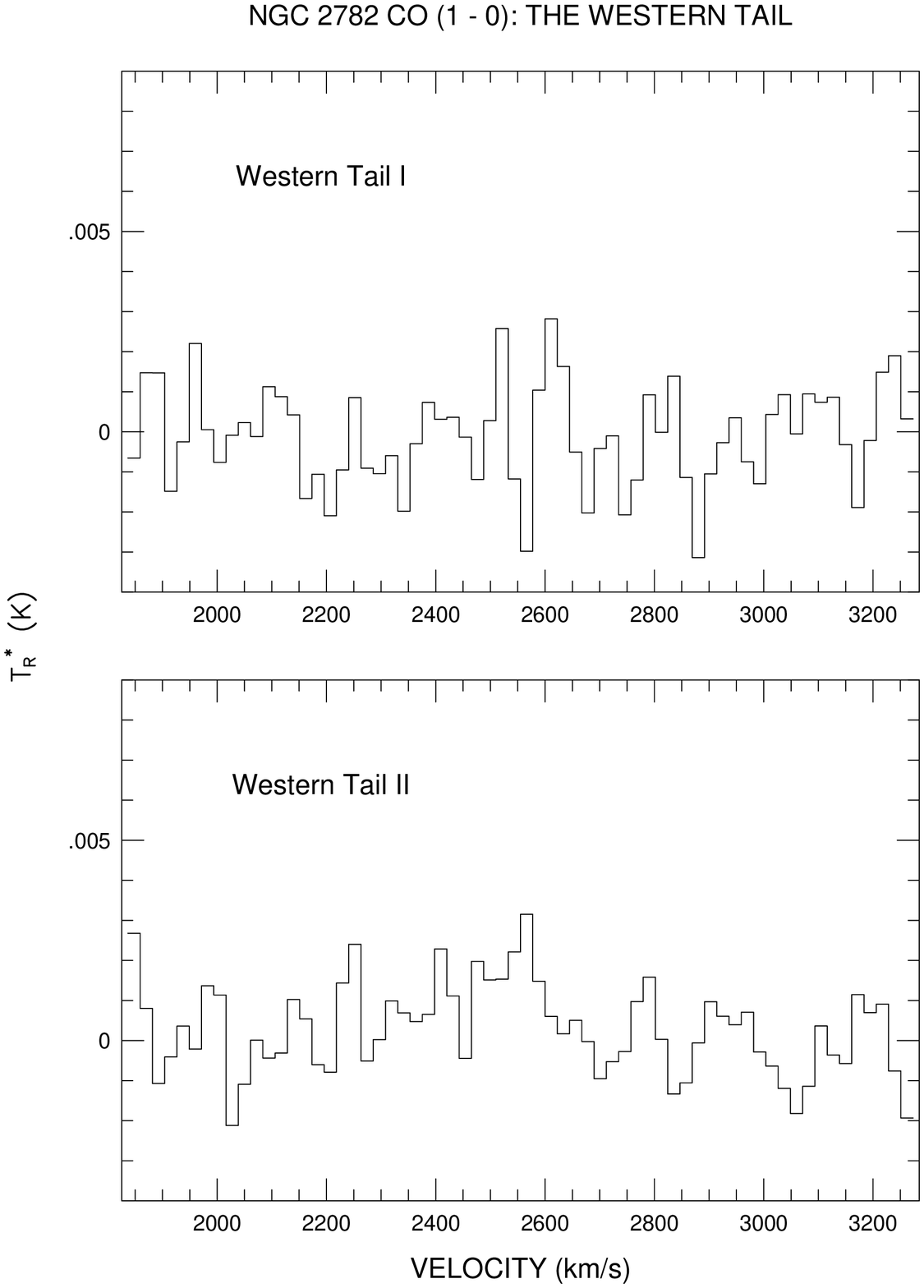]{The CO spectra for the observed positions in
the western tail.  
These data have been smoothed and resampled as in Figure 2.
Position I is the more southern position in the western tail.}

\figcaption[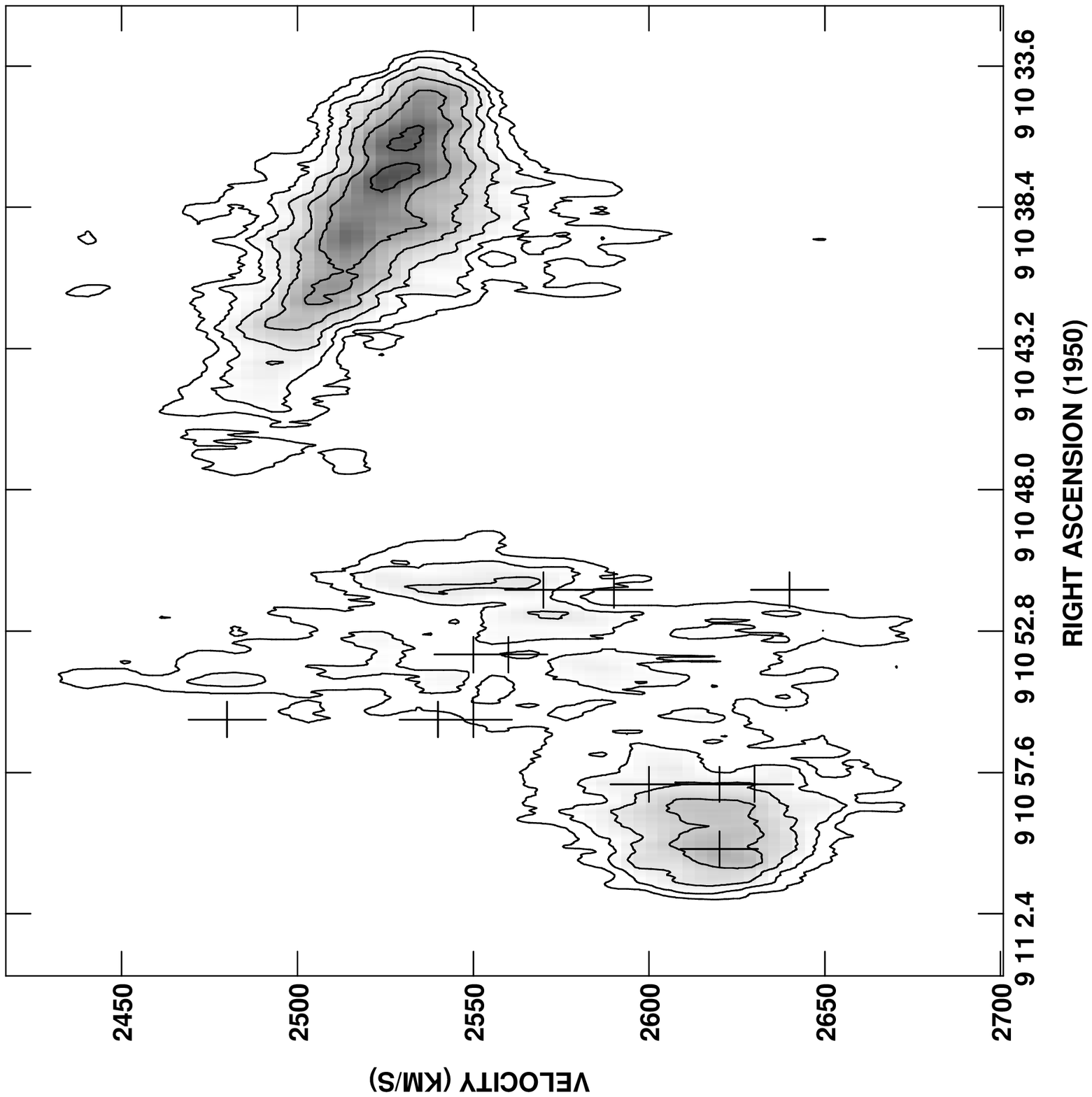]{A right ascension-velocity map for NGC 2782.
The contours and greyscale are the 21 cm HI data from Smith (1994).
The crosses mark the central velocities of the 12m CO lines.
The HI data have been averaged over declination.
The redshifted structure at 9$^{\rm h}$ 10$^{\rm m}$ 58$^{\rm s}$ is the eastern tail;
the blueshifted feature at 9$^{\rm h}$ 10$^{\rm m}$ 40$^{\rm s}$ is the western tail.
The disk lies 
at 9$^{\rm h}$ 10$^{\rm m}$ 53$^{\rm s}$, and is blueshifted to the east and 
redshifted to the west. }

\figcaption[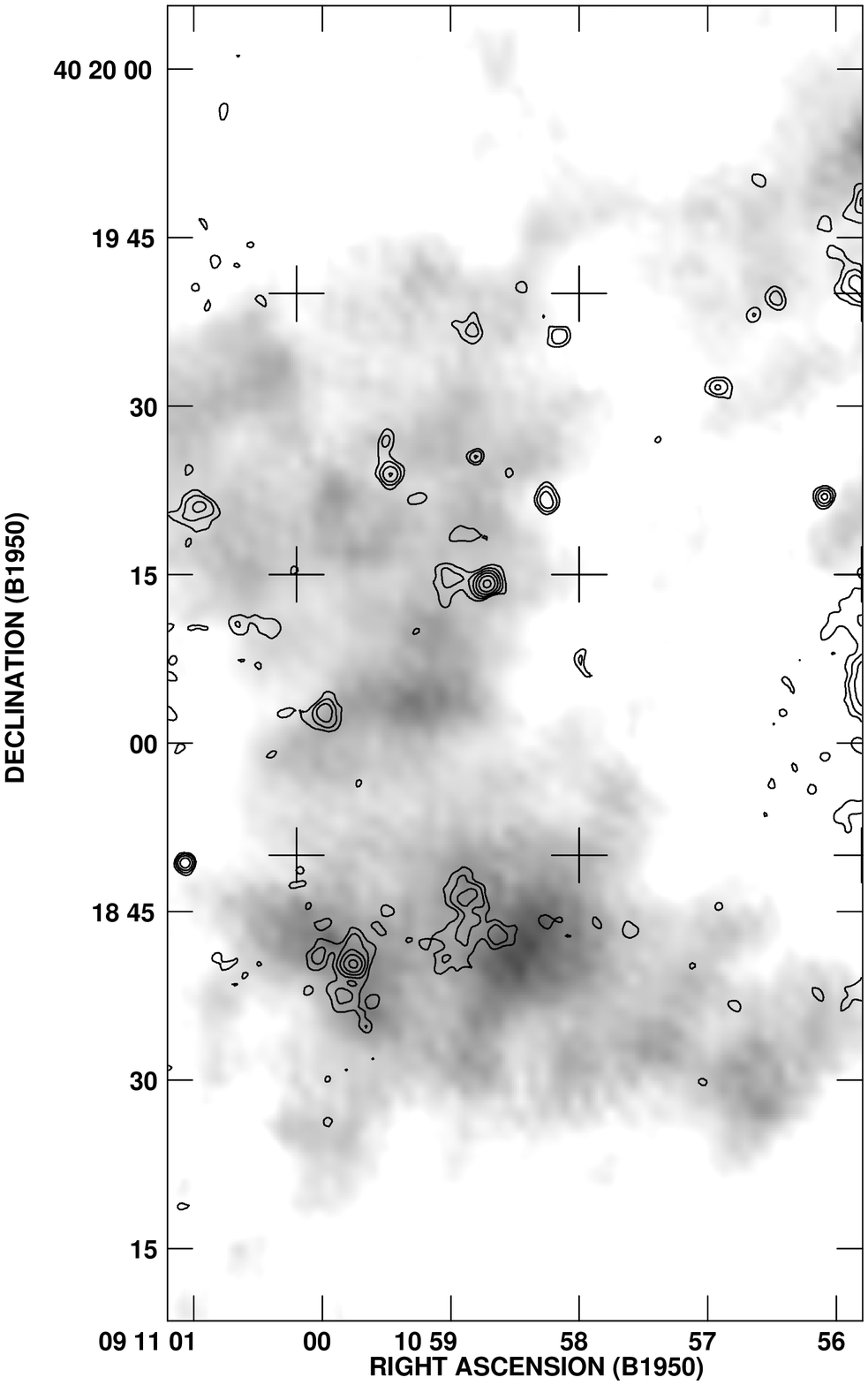]{The 21 cm HI map of NGC 2782 (greyscale; from Smith 1994),
overlaid with H$\alpha$ contours (from Jogee et al. 1998).
The H$\alpha$ map has been smoothed by a 1$''$ Gaussian to emphasize
the faint H~II regions.  The first contour level corresponds to
log F$_{H\alpha}$ = $-$16.5,
where F$_{H\alpha}$ is in units of erg s$^{-1}$ cm$^{-2}$ arcsec$^{-2}$.
The contour interval is log F$_{H\alpha}$ = 0.2.
The crosses mark the locations of our CO beams.
The feature at 9$^{\rm h}$ 10$^{\rm m}$ 58$^{\rm s}$, 40$^{\circ}$
19$'$ 8$''$ is likely an artifact from imperfect stellar subtraction.}


\begin{references}

\reference{a96}Arimoto, N., Sofue, Y., $\&$ Tsuijimoto, T. 1996, PASJ,
48, 275

\reference{a66}Arp, H. C. 1966, Atlas of Pecular Galaxies
(Pasadena: California Institute of Technology)

\reference{b83}Balzano, V. A. 1983, ApJ, 268, 602

\reference{b88}Barnes, J. E. 1988, ApJ, 331, 699

\reference{bbd90}Blitz, L., 
Bazell, D., $\&$ D\'esert, F. X. 1990, ApJ, 352, L13

\reference{b86}Bloemen, J. B. G. M., et al. 1986, A$\&$A, 154, 25

\reference{bsk92}Boer, B., Schultz, H. $\&$ Keel, W. C. 1992, A$\&$A, 154, 25

\reference{bhb92}Brouillet, N., Henkel, C., $\&$ Baudry, A.
1992, A$\&$A, 262, L5

\reference{c90}Cayatte, V., van Gorkom, J. H., Balkowski, C., $\&$ Kotanyi, C. 1990,
AJ, 100, 604

\reference{d88}Cohen, R. S., Dame, T. M., Garay, G., Montani, Rubio, M.,
$\&$ Thaddeus, P. 1988, ApJ, 331, L95

\reference{c85}Combes, F. 1985, Star Forming Dwarf Galaxies
and Related Objects, edited by D. Kunth,
T. X. Thuan, and J. Tran Thanh Van (Kim Hup Lee,
Singapore), 307

\reference{c88}Combes, F., Dupraz, C., Casoli, F., $\&$ Pagani, L. 1988, A$\&$A,
203, L9

\reference{dh89}Dettmar, R.-J., $\&$ Heithansen, A. 1989, ApJ, 344, L61

\reference{dbwm97}Duc, P.-A., Brinks, E., Wink, J. E., $\&$ Mirabel, I. F.
1997, A$\&$A, 326, 537

\reference{dm94}Duc, P.-A., $\&$ Mirabel, I. F. 1994, A$\&$A, 289, 83.

\reference{d84}Dufour, R. J. 1984, in IAU Symposium 108, Structure
and Evolution of the Magellanic Clouds, ed. S. van den Bergh and K. S.
de Boer (Dordrecht: Reidel), p. 353

\reference{e89}Elmegreen, B. G. 1989, ApJ, 338, 178

\reference{e95}Elmegreen, D. M., Kaufman, M.,
Brinks, E., Elmegreen, B. G.,
$\&$ Sundin, M. 1995, ApJ, 453, 100

\reference{e96}Evans, I. N., Koratkar, A. P., Storchi-Bergmann,
T., Kirkpatrick, H., Heckman, T. M., $\&$ Wilson, S. A. 1996,
ApJS, 185, 93

\reference{f67}Faulkner, D. J. 1967, MNRAS, 135, 401

\reference{f79}Federman, S. R., Glassgold, A. E., $\&$ Kwan, J. 1979, ApJ, 227, 466

\reference{f94}Freedman, W. L., et al. 1994, ApJ, 435, L31

\reference{f80}French, H. B. 1980, ApJ, 240, 41

\reference{g97}Garc\'ia-Vargas, M. L., Gonz\'alez-Delgado, R. M.,
P\'erez, E., Alloin, D., D\'iaz, A., $\&$ Terlevich, E. 1997,
ApJ, 478, 112

\reference{g95}Gonz\'alez-Delgado, 
R. M., P\'erez, E., D\'ias, A. I., Garc\'ia-Vargas, M. L.,
Terlevich, E., $\&$ Vilchez, J. M. 1995, ApJ, 439, 604

\reference{h87}Harwit, M., Houck, J. R., Soifer, B. T., $\&$ Palumbo, G. G. C.
1987, ApJ, 315, 28

\reference{h95}Hibbard, J. E. 1995, Ph. D. Thesis, Columbia University

\reference{hm95}Hibbard, J. E., $\&$ Mihos, J. C. 1995, AJ, 110, 140

\reference{hv96}Hibbard, J. E., $\&$ van Gorkom, J. H. 1996, AJ, 111, 655.

\reference{hlk89}Hodge, P. W., Lee, M. G., $\&$ Kennicutt, R. C., Jr. 1989, PASP, 101, 32

\reference{hk83}Hodge, P. W., $\&$ Kennicutt, R. C. 1983, AJ, 88, 296

\reference{hm89}Hollenbach, D., $\&$ McKee, C. F. 1989, ApJ, 342, 306

\reference{hg85}Hunter, D. A., $\&$ Gallagher, J. S. 1985, ApJS, 58, 533

\reference{hhg93}Hunter, D. A., Hawley, W. N., $\&$ Gallagher, J. S. 1993, AJ, 106, 1797

\reference{i94}Irwin, J. A. 1994, ApJ, 429, 618

\reference{id91}Israel, F. P., $\&$ de Graauw, Th. 1991, 
in IAU Symposium 148, 
The Magellanic Clouds, ed. R. Haynes and 
D. Milne (Boston: Kluwer), p. 45

\reference{itb95}Israel, F. P., Tacconi, L. J., $\&$ Baas, F. 1995, A$\&$A, 295, 599

\reference{i97}Israel, F. P. 1997, A$\&$A, 317, 65

\reference{j92}Jacoby, G. H., et al. 1992, PASP, 104, 599

\reference{j98a}Jogee, S., Kenney, J. D. P., $\&$ Smith, B. J. 1998,
ApJ, 494, L185

\reference{j99}Jogee, S., Kenney, J. D. P., $\&$ Smith, B. J. 1999,
ApJ, in preparation

\reference{k97}Kaufman, M., Brinks, E., Elmegreen, D. M.,
Thomasson, M., Elmegreen, B. G., Struck, C.,
$\&$ Klaric, M. 1997, AJ, 114, 2323

\reference{k95}Kenney, J. D. P., Rubin, V. C., Planesas, P., $\&$ Young, J. S.
1995, ApJ, 438, 135

\reference{k91}Kenney, J. D. P., Scoville, N. Z., $\&$ Wilson, C. D.
1991, ApJ, 366, 432

\reference{k83}Kennicutt, R. C. 1983, ApJ, 272, 54

\reference{kennicutt84}Kennicutt, R. C. 1984, ApJ, 287, 116

\reference{k89}Kennicutt, R. C. 1989, ApJ, 344, 685

\reference{l88}Lada, C. L., Margulis, M., Sofue, Y., Nakai, N.,
$\&$ Handa, T. 1988, ApJ, 328, 143

\reference{m97}Madden, S. C., Poglitsch, A., Geis, N., Stacey, G. J., $\&$ Townes, C. H.
1997, ApJ, 483, 200

\reference{m90}Maloney, P. 1990, in The Interstellar Medium in Galaxies,
edited by H. Thronson and M. Shull (Kluwer Academic Press, Boston),
p. 493

\reference{mw96}Maloney, P., $\&$ Wolfire, M. G. 1996, IAU Symposium 170, CO:
Twenty-Five Years of Millimeter-Wave Spectroscopy (Tucson, AZ)

\reference{mb88}Maloney, P., $\&$ Black, J. H. 1988, ApJ, 325, 389

\reference{mbr93}Mihos, J. C.,
Bothun, G. D., $\&$ Richstone, D. O.
1993, AJ, 418, 82

\reference{m91}Mirabel, I. F. et al. 1991, A$\&$A, 243, 367

\reference{m92}Mirabel, I. F. et al. 1992, A$\&$A, 256, L19

\reference{m94}Mochizuki, K. et al. 1994, ApJ, 430, L37

\reference{mv94}Morris, S. L., $\&$ van den Bergh, S. 1994, ApJ, 427, 696

\reference{p95}Poglitsch, A., Krabbe, A., Madden, S. C., Nikola, T.,
Geis, N., Johansson, L. E. B., Stacey, G. J., $\&$ Sternberg, A. 1995, ApJ,
454, 293

\reference{r91}Rubio, M., Garay, G., Montani, J., $\&$ Thaddeus, P.
1991, ApJ, 368, 173

\reference{r93}Rubio, M., et al. 1993a, A$\&$A, 271, 1

\reference{rlb93}Rubio, M., Lequeux, J., $\&$ Boulanger, F. 1993b, A$\&$A, 271, 9

\reference{rd90}Russell, S. C., $\&$ Dopita, M. A. 1990, ApJS, 74, 93

\reference{sow73}Sakka, K., Oka, S., $\&$ Wakamatsu, K. 1973, PASJ, 25, 153

\reference{sb94}Sandage, A., $\&$ Bedke, J. 1994,
The Carnegie Atlas of Galaxies (Washington D.C.: Carnegie
Institute)

\reference{s77}Savage, B. D., Drake, J. F.,
Budich, W., 
$\&$ Bohlin, R. C.
1977, ApJ, 216, 291

\reference{s78}Schweizer, F. in Structure and Properties of Nearby Galaxies, 1978,
279.

\reference{s82}Schweizer, F. 1982, ApJ, 252, 455

\reference{s83}Shaver, P. A., McGee, R. X., Newton, L. M., Danks, A. C.,
$\&$ Pottash, 1983, MNRAS, 204, 53

\reference{s91}Smith, B. J. 1991, ApJ, 378, 39

\reference{s94}Smith, B. J. 1994, AJ,  107, 1695

\reference{s97}Smith, B. J. 1997, AJ, 114, 2177

\reference{sh94}Smith, B. J., $\&$ Higdon, J. L. 1994, AJ, 108, 837

\reference{ssp97}Smith, B. J., Struck, C., $\&$ Pogge, R. W. 1997, ApJ, 483, 754

\reference{smith98}Smith, B. J., et al. 1998, in preparation

\reference{sb51}Spitzer, L, Jr., $\&$ Baade, W. 1951, ApJ, 113, 413

\reference{str97}Struck, C. 1997, ApJS, 113, 269

\reference{s98}Struck, C. et al. 1998, in preparation.

\reference{ty86}Tacconi, L. J., $\&$ Young, J. S. 1986, ApJ, 308, 600

\reference{ty87}Tacconi, L. J., $\&$ Young, J. S. 1987, ApJ, 322, 681

\reference{t64}Toomre, A. 1964, ApJ, 139, 1217

\reference{tt72}Toomre, A., $\&$ Toomre, J. 1972, ApJ, 178, 623

\reference{v79}van der Hulst, J. M. 1979, AJ, 71, 131

\reference{vmb94}van der Hulst, J. M., Mahoney, J. H., $\&$ Burke, B. F. 1994, preprint

\reference{vb88}van Dishoeck, E. F., $\&$ Black, J. H. 1988, ApJ, 334, 771

\reference{w90}Wallin, J. F. 1990, AJ, 100, 1477

\reference{w94}Wilson, C. D. 1994, ApJ, 434, L11

\reference{w95}Wilson, C. D. 1995, ApJ, 448, L97

\reference{w84}Wevers, B. M. H. R., Appleton, P. N., Davies, R. D.,
$\&$ Hart, L. 1984, A$\&$A, 140, 125

\reference{y96}Young, J. S., Allen, L., Kenney, J. D. P., Lesser, A.,
$\&$ Rownd, B. 1996, AJ, 112, 1903

\reference{yk89}Young, J. S., $\&$ Knezek, P. M. 1989, ApJ, 366, L11

\reference{yts83}Young, J. S., Tacconi, L. J., $\&$ Scoville, N. Z.
1983, ApJ, 269, 136

\reference{ytm94}Yoshida, M., Taniguchi, Y., $\&$ Murayama, T. 1994,
PASJ, 46, L195

\reference{ytm98}Yoshida, M., Taniguchi, Y., $\&$ Murayama, T. 1998,
AJ, submitted

\end{references}
\end{document}